\documentclass[letterpaper]{sig-alt-hotnets}
\usepackage[bookmarks,colorlinks=true,citecolor=black,urlcolor=black,linkcolor=black]{hyperref}
\usepackage{color}
\usepackage{xspace}

\newcommand{\eg}{{\em e.g., }}
\newcommand{\ie}{{\em i.e., }}

\newcommand{\yanc}{\textit{yanc}\xspace}

\begin{document}

%
%

	\conferenceinfo{Hotnets '13,}{
		November 21--22, 2013, College Park, MD, USA.}
	\CopyrightYear{2013}
	\crdata{978-1-4503-2596-7}
	\clubpenalty=10000
	\widowpenalty = 10000

	\title{
		Applying Operating System Principles to SDN Controller Design
	}

	\author{
		Matthew Monaco,
		Oliver Michel,
		Eric Keller
		\\
		University of Colorado at Boulder \\
		\texttt{\{first.last\}@colorado.edu}
	}

	\maketitle

	\begin{abstract}
		Rather than creating yet another network controller which 
provides a framework in a specific (potentially new) programming language
and runs as a monolithic application, in this paper we extend an existing
operating system and leverage its software ecosystem in order to serve
as a practical SDN controller.  
This paper introduces \emph{yanc}, a controller platform for software-defined networks which
exposes the network configuration and state as a file system, enabling user
and system applications to interact through standard file I/O,
and to easily take advantage of the tools available on
the host operating system. In \emph{yanc}, network applications are separate processes,
are provided by multiple sources, and may be written in any language.
Applications benefit from common and powerful technologies such as the virtual
file system (VFS) layer, which we leverage to layer a distributed file system
on top of, and Linux namespaces, which we use to isolate applications with 
different views (\eg slices). 
In this paper we present the goals and design of 
\emph{yanc}.  Our initial prototype is built with the FUSE 
file system in user space on Linux and has been demonstrated with a simple
static flow pusher application.  Effectively, we are making 
Linux the network operating system.

	\end{abstract}


%
%

	\newpage 
\section{Introduction}
\label{sec:intro}

The introduction of software-defined networks has generated tremendous buzz in
the past few years as it promises to ease many of the network management
head-aches that have been plaguing network operators for years~\cite{onfspec, 4d}.
Software-defined networking uses a
logically centralized control plane to manage a collection of packet processing
and forwarding nodes in the data plane.  It has been proposed that this requires
an operating system for networks~\cite{nox} which provides an interface to
program the entire network.  Applications on top of the operating system perform
the various management tasks by utilizing the operating system's interface. At a
high level, an OS manages the interface with hardware (network devices) and
allows applications (network management tasks) to run.

Despite a useful analogy, the practical realization is that  while extensible,
the current SDN controllers~\cite{nox, Floodlight, Ryu, Nettle, Maestro} are geared
towards single, monolithic network applications where developers can write
modules in the supported language using the API provided by the framework,
compile the entire platform, and run as a single process. 
An alternate approach is to use new languages and compilers that allow programmers
to specify the application with a domain specific language and run the compiled 
executable, still as a monolithic application~\cite{frenetic, pyretic}. 
Among the downsides of a monolithic
framework is that a bug in any part of the application (core logic, a
module, etc.) can have dire consequences on the entire system. 

Moreover, each of the existing controllers end up independently needing and developing a similar
set of required features.  
This results in a fragmented effort implementing common features where the main
distinguishing aspect in each case is commonly the language in which applications
are allowed to be written (\eg NOX--C++, Ryu--Python, Floodlight--Java,
Nettle--Has-kell, etc).  Further, these controllers are missing important
features like the ability to run across multiple machines (\ie a distributed
controller) -- limited to a hot standby (in the case of Floodlight) or a custom
integration into a particular controller (in the case of Onix~\cite{onix} on top
of NOX).  Even support for the latest protocol is lacking; many have yet to move
past OpenFlow 1.0 for which newer versions have been released, the latest
being 1.3.1~\cite{onfspec} --- even Floodlight, a commercially available
controller, only supports 1.0~\cite{bigswitchcontroller}.


In this paper, we explore the question: Is a network operating system
fundamentally that different from an operating system such that it requires a
completely new (network) operating system?  We argue that instead of building
custom SDN controllers, we should leverage existing operating system
technology in building an SDN controller.  
Our goal is to extend an existing operating system (Linux) and its user space
software ecosystem in order to serve as a practical network OS.

\subsection{Yanc}

\begin{figure}
  \centering
  \includegraphics[width=0.85\columnwidth]{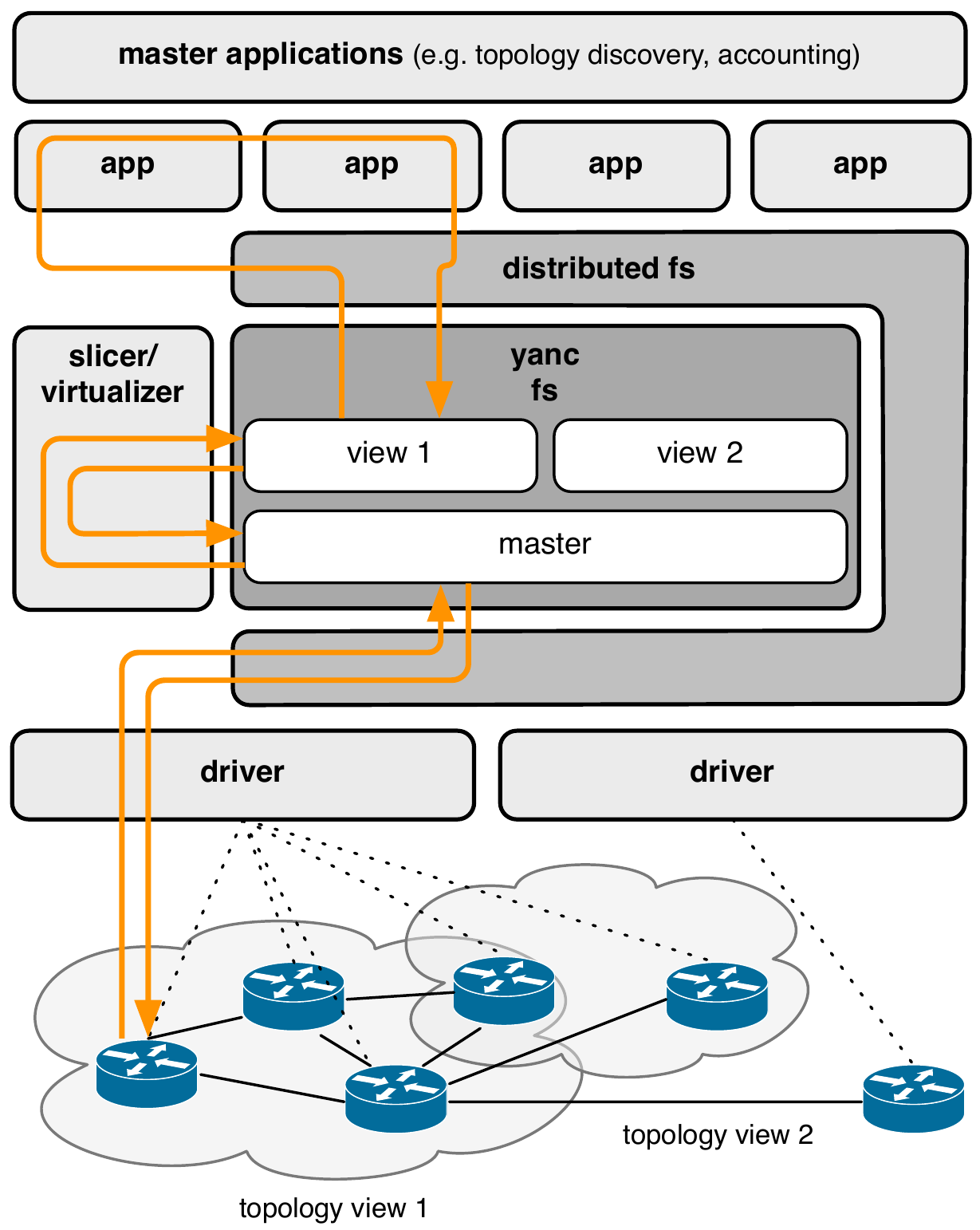}
  \caption{System Architecture}
  \label{fig:arch}
\end{figure}

We present the initial design of \yanc\footnote{\yanc, or yet another network
controller.}, 
which effectively makes Linux the network operating system and is 
rooted in UNIX philosophies (\S \ref{sec:goals}).  The \yanc
architecture, illustrated in Figure~\ref{fig:arch}, builds off of a central abstraction
used in operating systems today -- the file system.  With \yanc, the
configuration and state of the network is exposed as file I/O (\S \ref{sec:filesystem}) --
allowing running application software in a variety of forms (user space process, 
cron job, command line utility, etc) and developing in any language.  
Much like modern operating systems, system services interact with the real hardware
through drivers, and supporting applications can provide features such as virtualization,
or supporting libraries such as topology discovery (\S \ref{sec:apps}).  By using
Linux, we can leverage the ecosystem of software that has been developed for
it (\S \ref{sec:tech}).
One special example that is made possible by building \emph{yanc} into an existing operating
system is that distributed file systems can be layered on top of the \yanc file system
to realize a distributed controller (\S \ref{sec:dist}).  Finally, while we mostly discuss \yanc
in terms related to the OpenFlow protocol for ease of understanding, we believe the 
design of \yanc, extends into more recent research,
going beyond OpenFlow (\S \ref{sec:beyondof}).

We believe that once fully implemented,
\emph{yanc} will enable researchers to focus on value-added applications instead of yet
another network controller.

	\section{Goals}
\label{sec:goals}

Our goal is to extend an existing operating system (Linux) and its user space
software ecosystem in order to serve as a practical network OS. Hence, our
goals are similar to and extend from common Unix principles.

\textbf{Applications should encompass logically distinct tasks.} 
Each application should \textit{do one thing and do it well}.  For example,
there should be a distinct application for each protocol the network needs to
support such as DHCP, ARP, and LLDP. Furthermore, tasks such as logging, slicing,
and auditing should be packaged separately, likely with their own configuration
files.


\textbf{Applications may be written in any language.}
Rather than limiting applications to be written in the language in which the
controller is implemented or has provided bindings, developers should be able to
express logic in any language supported by modern operating systems.  This is in
contrast to the current approach where the decision for what controller to use
is greatly influenced by language of choice -- NOX (C++), Ryu (Python),
Floodlight (Java), Nettle (Haskell), etc.

\textbf{Applications should come from multiple sou\-rces.}
No one party will be an expert on all aspects of the network. Each application
should assume it will be working on a network with multiple black-box
applications. It is okay for one application to have specific dependencies on
others, but the dependency should be on a stable API. 
The file system API, for example, can be used heavily for exposing and
manipulating network state. For event driven applications that rely on timers
and other events, Linux already provides the necessary mechanisms.

\textbf{Applications should be decoupled from hardware.} 
A device driver is the implementation of a control plane protocol (\eg OpenFlow), or
even a specific version of a protocol (\eg OpenFlow 1.1). Drivers translate network
activity for a subset of nodes to the common API supported by the network
operating system. Nodes in such a system can therefore be gradually upgraded,
live, to newer protocols. Another implication of this is that multiple protocols
may be used simultaneously. 

\textbf{The interaction between applications should be defined by the administrator.}
Much of the time, multiple applications will be interested in the same network
events. Which applications have priority in responding to such events should be
up to the administrator of the network. Furthermore, the interaction between
applications should be flexible and not limited to a linear pipeline.

\textbf{Network application design should not be limitted by the controller.}
For some applications such as ARP and DHCP, a daemon processes is most
appropriate as they need to 
be ready to handle
messages at any time. Other applications, however, would be better designed as
simpler programs which are run occasionally. For example an auditor might run
periodically via a \texttt{cron} job, or  LIME \cite{lime}, will require the
occasional reshuffling of flow entries and is best called on-demand.

Furthermore, by exposing the network through a filesystem, small shell scripts
or even ad-hoc commands may be used to perform tasks on the network using
existing and well-known applications in the typical \texttt{coreutils} suite
such as \texttt{cat} and \texttt{grep}.

\section{The Yanc File System}
\label{sec:filesystem}

Central to \emph{yanc} is exposing network configuration and state as
a file system.  This decision stems from the fact that 
file systems are central to modern operating systems and enabling interaction
with network configuration and state through file I/O enables a powerful
environment for network administration. This follows from the file system abstraction
providing a common interface to a wide range of hardware,
system state, remote hosts, and applications. On Linux, ext4 controls access to
block devices, procfs to kernel state and configuration, sysfs to hardware, and
nfs to remote file systems.


In this section we describe how network configuration and state can be represented
with a file system abstraction.  We use aspects common to current SDN controllers
and protocols (namely switches, flows, etc.).   While we do have an initial prototype,
some details are yet to be worked out.

\subsection{File System Layout}

\begin{figure}
	\includegraphics{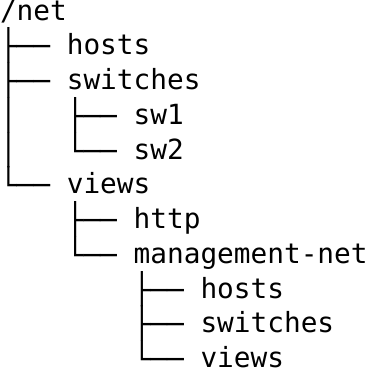}
	\caption{The \yanc file system hierarchy.}
	\label{fig:hier}
\end{figure}

At a basic level, file systems contain files which contain some information,
and directories which put structure around the collection of files.
The \emph{yanc} file system, as shown in Figure \ref{fig:hier}, is typically mounted on
\texttt{/net} and uses a directory structure to mimic the hierarchical 
nature of network configuration and state.  Top level directories will represent 
coarse entities such as \texttt{switches/} and
\texttt{hosts/}, as well as alternate representations of the state, which we call
views (discussed in \S \ref{sec:apps}). 


The actual configuration (\eg port status) and state (\eg counters)
are represented through a combination of files and subdirectories. 
For example, a port can be brought down by
\begin{verbatim}
# echo 1 > port_2/config.port_down
\end{verbatim}

Importantly, with \yanc, directories and files contain semantic information.
With this, each directory which contains a list of objects automatically creates an object
of the appropriate type on a \texttt{mkdir()} or \texttt{create()} system call.
For example
\begin{verbatim}
# mkdir views/new_view
\end{verbatim}
will create the directory \texttt{new\_view}, but also the \texttt{hosts},
\texttt{switches}, and \texttt{views} subdirectories.

\subsection{Switches}

\begin{figure}
	\includegraphics{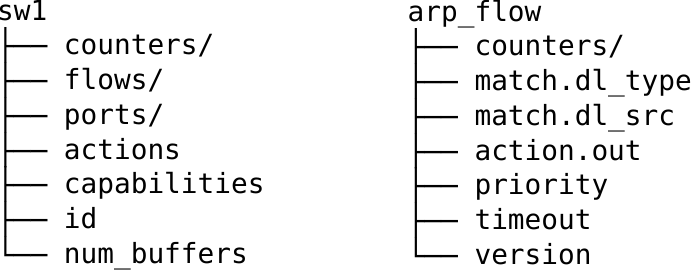}
	\caption{Partial representations of a \yanc switch and flow.}
	\label{fig:switch}
	\label{fig:flow}
\end{figure}

As mentioned, a switch is a coarse entity and each will be represented
with its own directory.
Our current switch representation is given in Figure \ref{fig:switch} -- with
files containing information about the switch (\eg capabilities) and 
directories pointing to further specific configuration and state (\eg ports). 
Switches can be created, deleted, and renamed with the standard file system calls
(\texttt{mkdir()}, \texttt{rmdir()}, and \texttt{rename()}, respectively).
Children of this object do not need to be removed prior to removing the object
itself; in other words, the \texttt{rmdir()} call for switches is automatically
recursive.

\subsection{Ports and Topology}

Each switch contains a set of ports.  Ports are much like switches in that
each is represented with a directory, and within the directory are
files with information about the given port.  This can include
feature sets (\eg what line rate is supported), state (\eg counters),
status, and configuration information (\eg hardware address).

To represent topological information, \emph{yanc} leverages symbolic links
for simpler manipulation or re-organization of directory structure
while still conveying the important link information (\eg rather than
parsing some topology information file).  
Each port contains a
symbolic link named \texttt{peer} which may or may not exist. Pointing this
link at another port is how physical links are represented. It is currently an
error to point this symbolic link at anything other than a port. Link status is
configured and inspected through the file \\ 
\texttt{config.port\_status}.

\subsection{Flows}

Within OpenFlow, and even many of the proposals for SDN protocols
beyond OpenFlow, a flow entry is a table entry which indicates
how the switch should handle traffic.  As flows contain many
pieces of information, each
\yanc flow (Figure \ref{fig:flow}) is represented by a directory.  Each field
that can be matched is a separate file \eg \texttt{match.nw\_src}.  Absence of a
match file implies a wildcard, and fields such as IP source take the CIDR
notation. Similarly, files named \texttt{action.*} are used for specifying
actions and files such as \texttt{priority} are used for setting other
properties of a flow entry.

In order for multiple values to be set atomically, we currently use a
\texttt{version} file. Applications use inotify or fanotify (\S
\ref{sec:fs-monitoring}) to monitor the \yanc file system for new, changing, or deleted
flow entries. Changes are only sent to hardware (by the drivers (\S
\ref{sec:drivers}) 
once the version has been incremented.

The flow written by one application may need to be modified by another before
being written to hardware. This can be for performance, security, or monitoring
purposes. We believe that the views abstraction (\S \ref{sec:views}) handles the
majority of these situations. However, composing networks from independent rules
is still an ongoing area of research \cite{pyretic}.

\subsection{Packet In}

Events originating on switches must be notified to applications.
With \yanc, events are also handled through file I/O.  Specific events
relating to the switch hardware status are handled by the drivers (\S
\ref{sec:drivers}).
For network management applications, the main event is the packet-in message 
in OpenFlow, though any event will be handled similarly.
Packet-in events are typically caused by table misses,
but also by explicit rules to trigger software processing. A simple example of
this is when software must respond to a new flow in a reactive network. 

Our current design concurrently feeds packet-in messages to all applications
interested in such events. Views (\S \ref{sec:views}) are used when one
application needs to alter a packet-in before it is received by another (\eg
to provide a virtual switch abstraction).

To be able to receive events, 
each application interested in packet-in events creates a directory in the
\texttt{events/} subdirectory (effectively creating a private buffer). New
packet-in messages appear as subdirectories within each private buffer, with 
files containing information about the message. 

	\section{System Applications}
\label{sec:apps}

The file system is effectively a mechanism to represent network configuration and state -- 
it cannot stand on its own.  With \yanc, system applications form the core of necessary
applications that every network management system will consist of.   In this section
we describe the drivers which interact with the physical switches, the
applications which alter the view of the network for other applications,
and the applications which discover the topological information about the network.

\subsection{Drivers}
\label{sec:drivers}

Analogous to device drivers in operating systems, 
device drivers in \yanc are a thin component which speaks the programming
protocol supported by a collection of switches in the network. Multiple drivers
for different protocols --- or even protocol versions --- are supported. For
example the majority of switches will communicate with an OpenFlow 1.0 driver, a
handful with a separate OpenFlow 1.3 driver, and others with a driver for an
experimental protocol being developed.  Applications will be able to 
inspect the files in the switches directory to know whether newer features
such as quality-of-service are supported or not.   With the file
system as the API, supporting new protocols only requires a new
driver to write new files, it does not require modifications to the 
core controller and interface provided to applications.  

\subsection{Network Views}
\label{sec:views}

Network virtualization has gone hand-in-hand with software-defined networking
since SDN's inception, and has proven to be one of the `killer applications'
of SDN~\cite{casadopresto2010}.  With \yanc, we term any logical representations of an 
underlying network a network view.  To create a new view, an application effectively
interacts with two portions of the file system simultaneously -- providing a translation
between them.  Then, any application that works off of a view will simply interact
with the associated portion of the file system (which can be isolated
with Linux namespaces as discussed in \S \ref{sec:namespaces}).
With \yanc, views can be stacked arbitrarily on top of
one another to facilitate any logical topology and federated control required of
the network.

The two main transformations in a network (today) are slicing and virtualization. A
slice of a network is a subset of the hardware and header space across one or
more switches, however the original topology is not changed \cite{flowvisor}.
Network virtualization on the other hand, provides any arbitrary transformation, 
such as combining multiple switches and forming a new
topology through the use of overlays. 
These two concepts can be combined to \eg slice traffic on port 22 out
of the network, and then create a virtual single-big-switch topology.



\subsection{Discovery and Topology}

Building a topology for other applications to use is essential to the function
of the control plane. As with everything in \yanc, topology is represented in the
file and directory layout of the file system. A topology application will handle
LLDP messages for discovery and create symbolic links which connect source to
destination ports. Other applications can make use of this topology to \eg build
network maps or calculate routes.

\section{Leveraging Existing Software}
\label{sec:tech}

The true benefit of \yanc comes from being able to use the file system in
conjunction with existing (Linux) technologies and software. Here, we discuss
some key components which are useful in developing a network control plane.

\subsection{Permissions, ACLs, and Attributes}

The VFS layer (\S \ref{sec:filesystem}) includes basic Unix permissions,
access control lists (ACLs), and extended attributes. Modern, multi-user
operating systems use permissions to control access to files and directories.
Likewise, the network operating system can implement fine-grained control of
network resources using permissions. For example, while individual flows can be
protected for specific processes, so too can an entire switch (thus all
of its flows).  Making use of file permissions for network control has benefits
in security, stability, and isolation.

Extended attributes are another form of metadata commonly found in file systems.
In \yanc, extended attributes can be used in arbitrary ways by developers. We
plan on utilizing them to specify consistency requirements for various network
resources (\S \ref{sec:dist}).

\subsection{File System Monitoring}
\label{sec:fs-monitoring}

Applications using the \yanc file system will use one of the Linux fsnotify APIs
--- inotify or fanotify --- to monitor for changes in the network. For example,
to monitor for new switches a watch can be placed on the \texttt{switches}
directory. To monitor for a changed flow, a watch can be placed on the
\texttt{version} file of a particular flow. This system fits in well with
modern, single-event-loop designs. Furthermore, use of the *notify systems comes
free, requiring no additional lines of code to the \yanc file system.

\subsection{Namespaces and Control Groups}
\label{sec:namespaces}

Linux namespaces allow, on a single system, isolation of resources such as
network interfaces, processes, users, and more. Control groups allow processes
to be grouped in an arbitrary hierarchy for the purpose of resource management
\eg CPU, memory, and disk IO usage.

These technologies play an important role in all but the most basic of SDN
deployments. They can be used to isolate subsets of the network to individual
processes, Linux containers such as OpenVZ, virtual machines, and even hosts
over the network. Users of these subsets are individuals, testers, federated
administrators, research groups, and data-center tenants.

\subsection{Standard Utilities}

As previously mentioned, we believe that interaction with the SDN devices should
be possible in many forms and not be relegated to sub-modules within a
monolithic application or some interface such as a REST API.  Nothing embodies
this more than the rich set of command line utilities available in modern
operating systems.

A quick overview of the switches in a network can be provided by:
\begin{verbatim}
$ ls -l /net/switches
\end{verbatim}

To list flow entries which affect ssh traffic:
\begin{verbatim}
$ find /net -name tp.dst -exec grep 22
\end{verbatim}

From simple one-liners to more elaborate shell scripts, these common utilities
are tools that system administrators use and know, thus they should be
able to be utilized in a software-defined network.

\section{Distributed Control}
\label{sec:dist}

When discussing SDN, papers are often careful to state that it uses a
\emph{logically centralized} controller, with the implication that multiple
servers can be used to realize a single unified network control.  To do so, each
monolithic controller application must duplicate effort and implement support for it.
Some example controllers include Onix, an extension of the NOX controller to
integrate support for distributed controllers; ONOS~\cite{onos} looks to have
much in common with Onix, but open source and based on Floodlight instead of
NOX; and Floodlight itself commercially supports a hot swap capability but not a
fully distributed controller.  In each case, they chose a distribution
technology that's forced on the user, and is tied to a specific controller.  

We argue that you can layer any number of distributed file systems on top of the
\yanc file system and arrive at a distributed SDN controller.  Each
distributed file system has a different implementation (centralized,
peer-to-peer with a DHT, etc.) with varying trade-offs. 
As a proof of concept, we mounted NFS on top of \yanc and
distributed computational workload among multiple machines. Other layered file systems
which are likely to be useful on top of \yanc are sshfs for secure, remote access
and WheelFS which provides distribution and configurable consistency in a
logically centralized hierarchy.

	\section{Beyond OpenFlow}
\label{sec:beyondof}

While much of the discussion has centered around switches and to some degree,
OpenFlow, \yanc is not tied to either.

\subsection{Network controller, or network device?}


With a distributed file system, the lines between controller and device start to become
blurred.  Similar to what was envisioned with Kandoo~\cite{kandoo}, we envision
network devices potentially running software that today is largely relegated to
the network controller on a remote server.  We would like to take advantage of
vendors such as Cumulus Networks\cite{cumulus} and Pica8\cite{pica8} which are
providing devices coupled with a customizable Linux operating system.

These devices can run \yanc and participate in a distributed file system rather
than have a bespoke communication protocol that's dependent on how the control
logic was partitioned.  With \yanc, when an application on another machine writes
to a file representing a flow entry, that will then show up on the device (since
it's a distributed file system), and the device can read it and push it into the
hardware tables.  Similarly, software running on a switch can make a change
locally and this will be seen by remote servers.
%
%
While unproven, we believe this will will be a fruitful direction to explore --
particularly in the quest to extend SDN to middleboxes.

\subsection{Extending to Middleboxes}

Given the limitation of OpenFlow to layers 2 through 4 with predefined processing,
there has been recent research activity in extending SDN principles into
middleboxes~\cite{sdmiddlebox, nfv, splitmerge, hotsdnclick}.  

For a middlebox with fixed funcationality, but exposing its state through 
a standardized protocol, a driver can be written to populate and interact
with the file system and take immediate advantage of \yanc -- such as what's 
involved in the move from OpenFlow 1.0 to 1.3.
This interface can be used to move the state around to elastically expand the
middlebox~\cite{splitmerge}.  We envision that we can use command line
utilities such as \texttt{cp} or \texttt{mv} to move state around rather than custom
protocols.  

More interesting is the case where the device allows for  custom processing.
To achieve this with OpenFlow, we can either
use packet-in events between device and controller or re-direct packets to
custom boxes (even directly attached~\cite{flowstream}).  
Using \yanc coupled with a distributed file
system, the device itself can run \yanc, run the application software, and still
be able to work under the direction of global network view.

	\section{Prototype}
\label{sec:proto}

We have built a prototype of \yanc which consists of a number of the components
described in this paper. Most importantly, the core file system is implemented
via the FUSE user space file system\cite{fuse} (which provides for easier
initial prototyping). The current \yanc file system supports the introspection and manipulation of network
elements, flows, ports, links, and sliced/virtual networks. Multiple tables and
queues are not yet implemented.  The prototype has initial but incomplete support
for properly handling table misses and flow table writes in a multiprocess
system.  


An OpenFlow driver is implemented in C++ and supports all of the protocol
features supported by the file system. A simple ``static flow pusher'' shell
script can be used to write flows to switches. A topology daemon is implemented
in Python and maintains port-to-port symbolic links. Finally, a router daemon
handles all table misses and sets up paths based on exact match through the
network.

\subsection{Performance}

The file system provides a flexible namespace, access control, and a familiar
programming pattern for network operators. However, it comes with some
performance cost.  Each fine-grained access to the file system is done through a
system call --- for example \texttt{read()}, \texttt{write()}, and
\texttt{stat()} --- which switches context from the application to the kernel.
Complex operations such as writing flow entries to thousands of nodes will
result in tens of thousands of context switches and thus a small performance
impact.

To mitigate the performance overhead of working with the file system, we are
implementing \texttt{libyanc}, a set of network-centric library calls atop a
shared memory system. The library provides a \textit{fastpath} for \eg creating
flow entries atomically and without any context switchings. It also allows for
the efficient, zero-copy passing of bulk data --- packet in buffers, for example
--- among applications

	\section{Conclusion and Future Work}
\label{sec:concl}

We have presented \yanc, a vision of how operating system mechanisms and
principles can be applied in the context of software-defined networking.
Effectively, \yanc realizes a network operating system which can be used in a
variety of ways in order to leverage innovation in the operating system space.
Thus, more focus can be put on specific control-plane-centric topics such as
load balancing, congestion control, and security. Our initial prototype is only
a first step toward realizing many of the aspects of \yanc that we introduced in
this paper.

%
%

	\footnotesize
	\bibliographystyle{abbrv}
	\bibliography{refs}

\end{document}